\documentclass[11pt,a4paper,notitlepage]{revtex4-1}
\usepackage{amsmath,amsfonts,amssymb,graphicx}
\usepackage{times}
\usepackage{bm}
\usepackage{lineno}

\begin{document}
\title{Light Nuclei and Hyper Nuclei Collectivity Measurements at High Baryon Density Region}

\author{Xionghong He}

\address{Institute of Modern Physics, Chinese Academy of Sciences, China}

\begin{abstract}
  High energy heavy-ion collisions produce large amounts of light nuclei and hype nuclei, especially at high baryon density around collision energy of several GeV.
	These light nuclei and hyper nuclei carry the information of nucleon-nucleon and hyperon-nucleon interactions and affect the chemical composition and properties of the collision system. 
	This proceeding is a brief review for the recent measurements on light nuclei collective flow, including directed flow $v_1$, elliptic flow $v_2$, and high order flow coefficients, at finite and high baryon densities from different experiments. The light nuclei production mechanism is discussed based on comparisons of the measurements and model calculations. The first measurement for hyper nuclei $v_1$ at 3 GeV by STAR is also reported, which may imply the hyper nuclei are formed via the coalescence of nucleons and hyperon.
\end{abstract}

\maketitle

\section{Introduction}
\label{intro}

The properties of strongly interacting matter under high temperatures and high densities are important research focuses of high energy heavy-ion collision experiments all over the world. It is still an open question whether the phase transition from hadronic matter to the quark-gluon plasma (QGP) is a first-order ~\cite{Bzdak_2020} and a critical point exists on the QCD phase diagram at finite baryon density~\cite{Nahrgang_2016}. 
Even if the QGP is not formed it is important to know the evolution of nuclear matter, which the thermodynamical properties are described by an equation-of-state (EoS).
The researches on the EoS can also shed new light on the nuclear matter state existing in the dense stellar objects given their high baryon density, such as neutron star~\cite{Most_2022}.
The azimuthal anisotropy in particle momentum distributions relative to the reaction plane, referred as collective flow, are thought be sensitive probes to the collision system properties at very stage of its evolution and to the EoS~\cite{Danielewicz_2002}. The collision energy dependence of the first- and second-order flow anisotropy parameters, $v_1$ and $v_2$, for different particle species provide valuable information on evolution of the nuclear matter. 
The high order flow parameters, $v_{n} (n>2)$ are usually originated from the initial state fluctuations which are not correlated to the reaction plane at very high collision energies~\cite{Adam_2016}.

The light nuclei flow has been measured by different experiments from low to high energies~\cite{Adamczyk_2016,Adam_2020,Abdallah_2022,Reisdorf_2012,Musch_2020,Wang_1995,Barrette_1999,Acharya_2017}. These measurements suggest that compared to protons which is the lighted nuclei, the light nuclei (deuteron, triton, $^{3}$He, and $^{4}$He) flow have more pronounced dependencies on collision energies, and thus may be more sensitive to the collective motion and EoS of nuclear matter. 
Moreover, the light nuclei flow is expected to reveal their production mechanisms in high energy heavy-ion collisions, which is still a question under debate. The statistical thermal model and nucleon coalescence model are two very popular but very different theoretical models for light nuclei production. 
The thermal model describes light nuclei production via the nucleon-nucleon or parton-parton interactions before the chemical freeze-out of the fireball~\cite{Mekjian_1978,Munzinger_1995}. It is however difficult to explain how light nuclei can survive in the hot temperature given their small binding energy.
The coalescence model describe the light nuclei is formed near kinetic freeze-out, when the system temperature is much lower, via combination of nucleons if these nucleons are near each other in coordinate and momentum space~\cite{Sato_1981,Steinheimer_2012}. One general feature of the nucleon coalescence model is that light nuclei flow is expected to follow an approximate atomic-mass-number ($A$) scaling under the assumption of small $v_{n}$
\begin{equation}
\label{eq-1}
v^{A}_{n}(p_{\rm T},y)/A \approx v^{p}_{n}(p_{\rm T}/A,y).
\end{equation}

If one of the nucleon in light nuclei is replaced by a hyperon, a hyper nuclei is formed. In heavy-ion collisions, the production of hyper nuclei allows study of hyperon-nucleon (Y$-$N) interaction, which remains unclear. At high baryon densities, the EoS is expected to be changed by the appearance of hyperons and hyper nuclei~\cite{Steinheimer_2012}, which is critical for understanding the inner structure of compact stellar objects~\cite{Gerstung_2020}. The thermal model calculations have predicted that the $^{3}_{\Lambda}$H and $^{4}_{\Lambda}$H production get maximum at high baryon density around several GeV~\cite{Andronic_2011}. The measurement of hyper nuclei flow provides valuable information for understanding the Y$-$N interaction as well as their production mechanism in heavy-ion collisions.

\section{Light Nuclei Flow}
\subsection{Light Nuclei $v_1$}
\label{sec-1}

\begin{figure}[htbp]
\centering
\includegraphics[width=9cm]{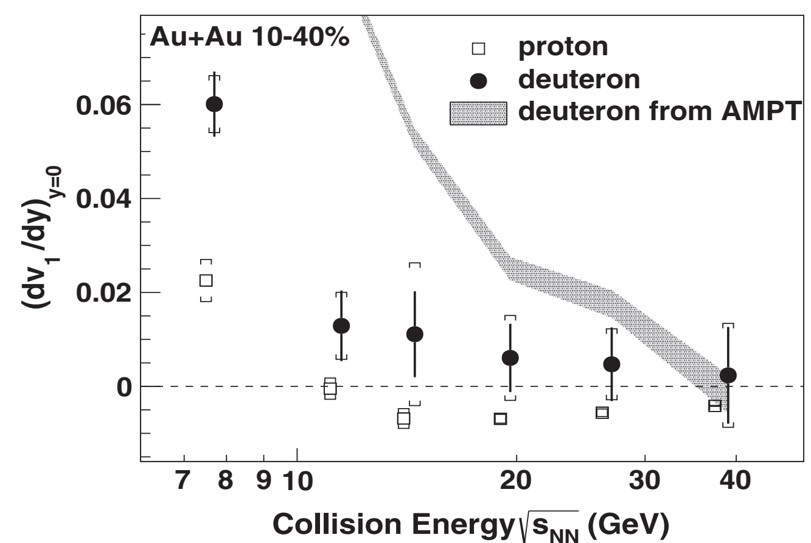}
\caption{The $v_1$ slopes at the midrapidity as a function of collision energy in 10-40\% Au+Au collisions from STAR experiment (figure from Ref.~\cite{Adam_2020}). The open circles and solid circles represent the results for deuterons and protons, respectively.}
\label{fig_1}      
\end{figure}

The energy dependence of $v_1$ for proton and net-proton at $\sqrt{s_{NN}}=$ 7.7 to 62.4 GeV have been measured by STAR experiment~\cite{Adamczyk_2014}.
The value of $v_1$ slopes at midrapidity, $dv_1/dy|_{y=0}$, change signs from positive to negative at $\sqrt{s_{NN}} \sim$ 10 GeV with the increased energy and show a minimum at $\sqrt{s_{NN}} \sim$ 15 GeV as shown in Fig.~\ref{fig_1}~\cite{Adam_2020}, which may be a signature of first-order phase transition 
according to a hydrodynamical model~\cite{Stocker_2005}. According to the picture of nucleon coalescence, the light nuclei $v_1$ should have a similar but more pronounced energy dependence behavior in the same energy region.

The deuteron $v_1$ has been measured by the STAR experiment and the extracted $dv_1/dy|_{y=0}$ as a function of collision energy is shown in Fig.~\ref{fig_1}~\cite{Adam_2020}. At $\sqrt{s_{NN}}=7.7$ GeV, the value of $dv_1/dy|_{y=0}$ for deuteron is higher than that for proton and follows the $A$ scaling considering the statistical and systematic uncertainties. 
Above 7.7 GeV, there is a hint that the deuteron $dv_1/dy|_{y=0}$ have a positive sign with large uncertainties, which is opposite to the corresponding proton $dv_1/dy|_{y=0}$. The $p_{\rm T}$ dependence of deuteron $v_1$ show an enhancement towards very low $p_{\rm T}$ at 7.7 GeV~\cite{Adam_2020}. This measurement may suggest a more complicated mechanism than the simple coalescence picture for light nuclei production, while a stronger conclusion needs more event statistics which can be achieved with the second phase of STAR beam energy scan program.

\begin{figure}[htbp]
\centering
\includegraphics[width=9cm,clip]{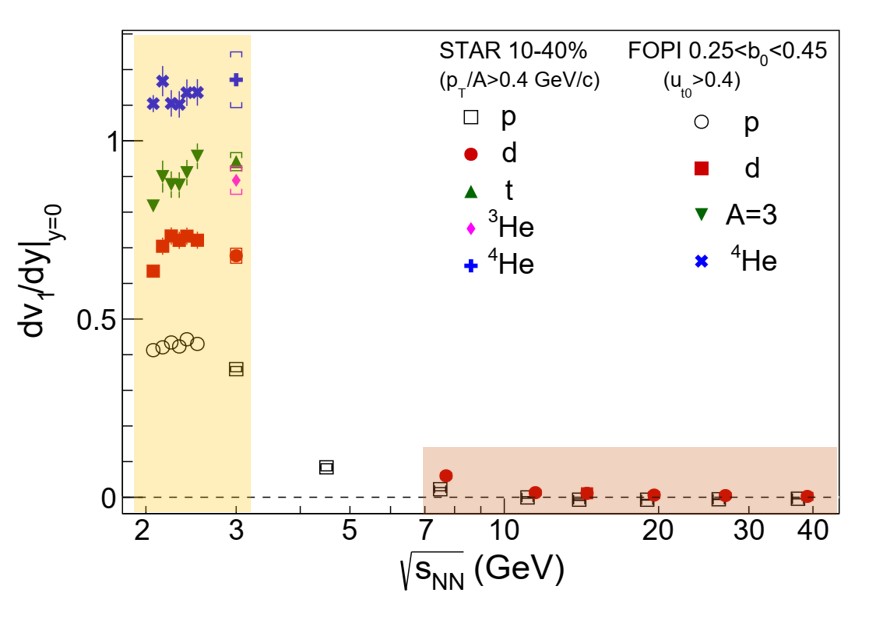}
\caption{The $v_1$ slopes at the midrapidity as a function of collision energy in 10-40\% Au+Au collisions from FOPI and STAR experiments~\cite{Adam_2020,Abdallah_2022,Reisdorf_2012}. The open markers represent proton and solid circles represent light nuclei. The results from STAR and FOPI are represented with different marker styles as they use different centrality definitions and transverse momentum cuts.}\label{fig_2}
\end{figure}

Moving to lower energies, the light nuclei $v_1$ slopes reach maximum and have a weak energy dependence from $\sqrt{s_{NN}}=$ 2 to 3 GeV~\cite{Adam_2020,Abdallah_2022,Reisdorf_2012}, as shown in Fig.~\ref{fig_2}. There is a clear mass ordering and the value of $dv_1/dy|_{y=0}$ follow the $A$ scaling approximately. 
Above 3 GeV, the $v_1$ slopes decrease with the increasing of collision energies. While the $dv_1/dy|_{y=0}$ has a steep drop with the decreasing energy below 2 GeV~\cite{Reisdorf_2012}.
The $p_{\rm T}$ dependencies of light nuclei $v_1$ also show the $A$ scaling at midrapidity~\cite{Abdallah_2022}. The scaling, however, worsens for higher $p_{\rm T}$ and away from midrapidity; one of possible reasons is the increasing contamination of target/beam rapidity fragments.
The proton $v_1$ at 3 GeV can be well reproduced by the baryon mean-field configuration of JAM transport model~\cite{Abdallah_2022}, while the cascade of the model underestimates the proton $v_1$. A nucleon coalescence using the proton and neutron phase space distributions from this JAM mean-field calculations can qualitatively describes the $v_1$ for all measured light nuclei species, which may suggest that the baryonic interactions dominate the evolution and light nuclei are likely formed via the coalescence of nucleons at energy of several GeV.

\subsection{Light Nuclei $v_2$}
\label{sec-2}
\begin{figure}[htbp]
\centering
\includegraphics[width=12cm,clip]{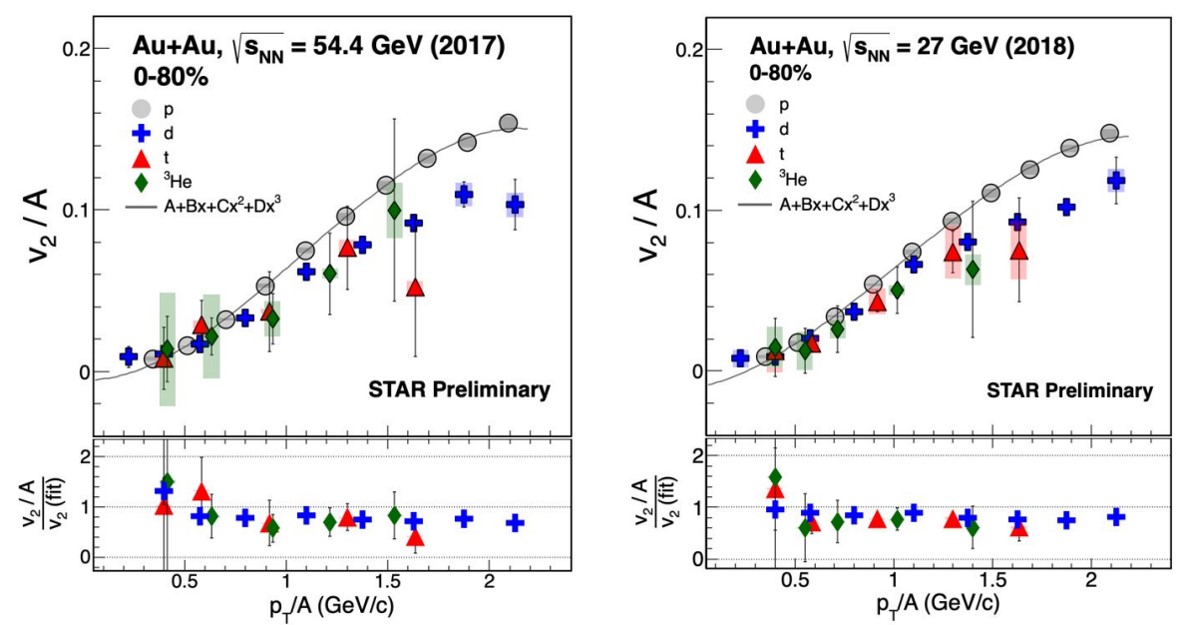}
\caption{The $v_2/A$ as a function of $p_{\rm T}/A$ in $\sqrt{s_{NN}}=$27 GeV and 54.4 GeV Au+Au collisions from STAR~\cite{qm22}. The bottom panels show the ratio of light nuclei $v_2$ to the fit of proton $v_2$.}
\label{fig_3}       
\end{figure}
The light nuclei $v_2$ as a function of $p_{\rm T}$ has been measured by the STAR experiment in $\sqrt{s_{NN}}=$ 7.7 to 200 GeV Au+Au collisions~\cite{Adamczyk_2016}. The $v_2$ value shows a monotonic rise with increasing $p_{\rm T}$ and a mass ordering with increasing mass number. The difference of $v_2$ between deuteron and anti-deuteron is qualitatively follow the difference between proton and anti-proton. For all measured collision energies, the $A$ number scaling holds generally for all (anti-)light nuclei at low $p_{\rm T}$ ($p_{\rm T}/A<1.5$ GeV/$c$), while the scaling is broken, especially for low energies, towards high $p_{\rm T}$. The results have been updated with more statistic events by STAR recently~\cite{qm22}, as shown in Fig.~\ref{fig_3}. 
Compared to $v_2$ of protons, the  $v_2/A$ have a overall 20-30\% deviations from $A$ scaling for deuteron, triton, and $^{3}$He at both $\sqrt{s_{NN}}=$ 27 GeV and $\sqrt{s_{NN}}=$ 54.4 GeV. The facts that both the deuteron $v_1$ and light nuclei $v_2$ do not have $A$ scaling imply that light nuclei may not formed only via a simple coalescence of nucleons at the energy region $\sqrt{s_{NN}}=7.7-200$ GeV.

\begin{figure}[htbp]
\centering
\includegraphics[width=9cm,clip]{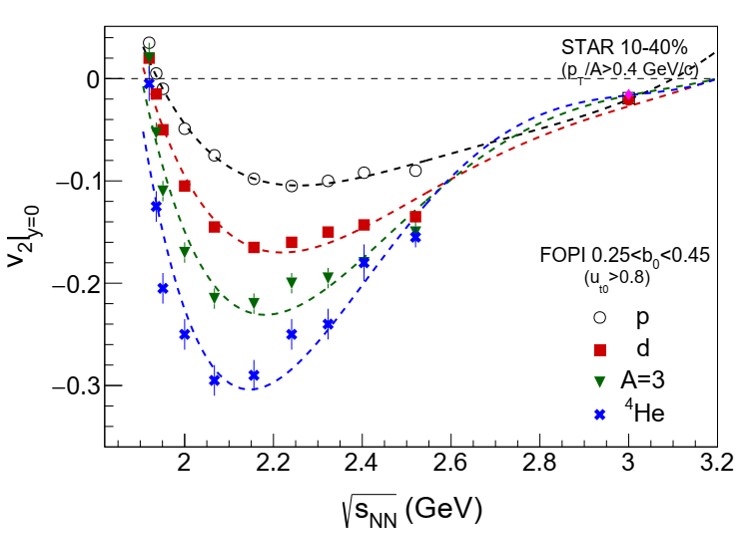}
\caption{The $v_2$ in the midrapidity as a function of collision energy. The data points are extracted from STAR and FOPI measurements~\cite{Abdallah_2022,Reisdorf_2012}. The dashed lines represent fits with polynomial functions to guide eyes.}
\label{fig_4}       
\end{figure}

With decreasing of collision energies, the $v_2$ value at midrapidity for protons as well as light nuclei become negative at energy close to 3 GeV, as in Fig.~\ref{fig_4}. 
This negative $v_2$ may be caused by the squeeze-out and/or shadowing of the spectators due to their passage time becoming comparable with the expansion time of the fireball at $\sqrt{s_{NN}}=$ 2 to 3 GeV~\cite{Reisdorf_2012}. The light nuclei $v_2$ at the mid-rapidity have a clear mass hierarchy that increases with increasing collision energy from 1.9 GeV to 2.2 GeV and then decreases from 2.2 GeV to 3 GeV.
The location of minimum value of $v_2$ varies with the nuclei mass. 
The mass hierarchy disappears at 3 GeV for all light nuclei. Similar to their $v_1$ slope behavior as a function of energy, the heavier nuclei have a stronger energy dependence which may suggest more sensitive to the EoS~\cite{Reisdorf_2012}. Although the light nuclei $v_2$ violate $A$ scaling at 3 GeV, the JAM plus coalescence calculations can qualitatively describe the rapidity dependence~\cite{Abdallah_2022}. The reason of the $A$ scaling broken may partly be that the Eq.~\ref{eq-1} is only valid under the assumption of small $v_1$~\cite{Abdallah_2022}. At 3 GeV, the $v_1$ is however not negligibly small as discussed in Section~\ref{sec-1}.

\subsection{Light nuclei $v_n(n>3)$}
HADES experiment measure the light nuclei flow from $v_1$ to $v_6$ using the first-order event plane at $\sqrt{s_{NN}}=2.4$ GeV Au+Au collisions. It is expected that high order flow coefficients ($n>3$) relative to the first-order event plane are zero if they originate from an initial state fluctuation. However, all the $v_3-v_6$ are not zero but have strong dependencies on particle $p_{\rm T}$ and rapidity for all the measured light nuclei species. No mass hierarchy is observed for $v_3-v_6$ in the midrapidity. Non-zero high order flow may imply these high order flows have more complicated causes at high baryon density and could be more sensitive to the EoS than $v_1$ and $v_2$ according to transport model calculations~\cite{Hillmann_2018}.

\section{Hyper nuclei $v_1$}
STAR experiment observed $v_1$ of $^{3}_{\Lambda}$H and $^{4}_{\Lambda}$H for the first time in $\sqrt{s_{NN}}=3$ GeV Au+Au collisions~\cite{sqm21}. The extracted $dv_1/dy|_{y=0}$ at the midrapidity are similar to the corresponding mass light nuclei within the statistical uncertainties. This measurement suggest that the two hyper nuclei may be formed via the coalescence of nucleons and $\Lambda$ hyperon.

\begin{figure}[htbp]
\centering
\includegraphics[width=9cm,clip]{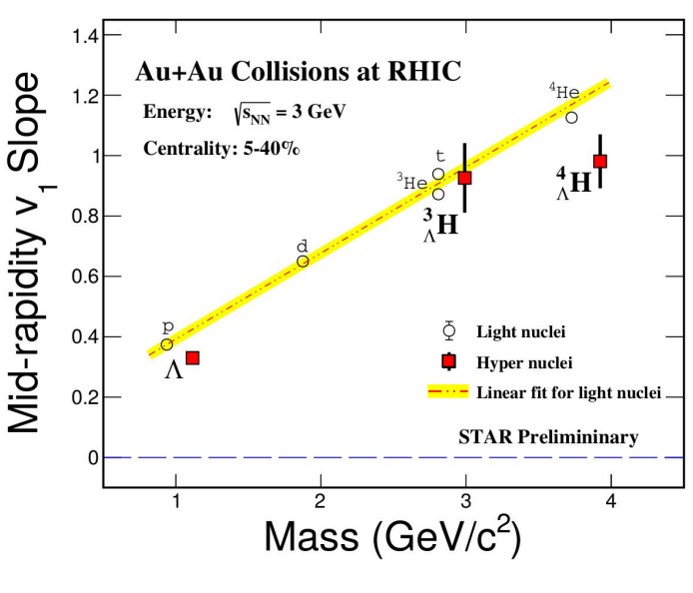}
\caption{The $v_1$ slope at the midrapidity as a function of baryon mass number in 5-40\% Au+Au collisions at $\sqrt{s_{NN}}=3$ GeV~\cite{sqm21}. The solid markers represent the $\Lambda$ baryon, $^{3}_{\Lambda}$H, and $^{4}_{\Lambda}$H. The open circles are the results for corresponding light nuclei. The dot-dashed line is a linear fit to the light nuclei $v_1$ slopes.}
\label{fig_5}       
\end{figure}

\section{Summary}
In summary, the light nuclei and hyper nuclei flow has been measured at finite and high baryon densities by several experiments. For collision energy above 7.7 GeV, there is a hint of opposite sign for deuteron $v_1$ slopes with the corresponding proton; the light nuclei $v_2$ show tensions with the expectations of coalescence model. The light nuclei $v_1$ slopes saturate and their $v_2$ at mid-rapidity are negative within 2$-$3 GeV. The data at 3 GeV can be qualitatively described by the coalescence model. Non-zero $v_3-v_6$ have been observed at 2.4 GeV which may be more sensitive to the EoS than $v_1$ and $v_2$ at high baryon densities. The first flow measurement for hyper nuclei suggests they may be formed via the coalescence of nucleons and hyperon in heavy-ion collisions.

\end{document}